\documentclass{aastex}
\usepackage{amssymb}
\usepackage{amsmath}
\usepackage{amsfonts}
\usepackage{graphicx}%
\begin{document}

\begin{center}
{\LARGE Influence of energy exchange of electrons and ions on the
long-wavelength thermal instability in magnetized astrophysical objects}\bigskip

{\LARGE \bigskip}{\large Anatoly K. Nekrasov}

Institute of Physics of the Earth, Russian Academy of Sciences, 123995 Moscow, Russia

anatoli.nekrassov@t-online.de, anekrasov@ifz.ru

\bigskip

{\large ABSTRACT}
\end{center}

We investigate thermal instability in an electron-ion magnetized plasma
relevant to galaxy clusters, solar corona, and other two-component
astrophysical objects. We apply the multicomponent plasma approach when the
dynamics of all the species are considered separately through electric field
perturbations. General expressions for perturbations obtained in this paper
can be applied for a wide range of multicomponent astrophysical and laboratory
plasmas also containing the neutrals, dust grains, and other species. We assume 
that background temperatures of electrons and ions are different and include the
energy exchange in thermal equations. We take into account the dependence of
collision frequency on density and temperature perturbations. The
cooling-heating functions are taken as different ones for electrons and ions.
As a specific case, we consider a condensation mode of thermal
instability of long-wavelength perturbations when the dynamical time is
smaller than a time during which the particles cover the wavelength along the
magnetic field due to thermal velocity. We derive a general
dispersion relation taking into account the effects mentioned above and
obtain simple expressions for growth rates in limiting cases.
Perturbations are shown to have an electromagnetic nature. We find
that at conditions under consideration transverse scale sizes of unstable
perturbations can have a wide spectrum relatively to longitudinal scale
sizes and, in particular, form very thin filaments. The results obtained
can be useful for interpretation of observations of dense cold regions in
astrophysical objects.


\bigskip

\section{INTRODUCTION}

The thermal instability leads to formation of regions with larger densities
and lower temperatures than that in the surrounding medium (Parker 1953; Field
1965). Beginning from the classical paper by Field (1965), this instability
was studied for both astrophysical objects (for reviews see, e.g.,
V\'{a}zquez-Semadeni et al. 2003; Elmegreen \& Scalo 2004; Cox 2005; Heiles \&
Crutcher 2005) and plasma physics applications (e.g., Meerson 1996). Majority
of papers were devoted to thermal instability in the interstellar medium
(ISM; e.g., Field 1965; Burkert \& Lin 2000; Hennebelle \& P\'{e}rault 2000;
Koyama \& Inutsuka 2002; Kritsuk \& Norman 2002; S\'{a}nchez-Salcedo et al.
2002; Audit \& Hennebelle 2005; Stiele et al. 2006; V\'{a}zquez-Semadeni et
al. 2006; Fukue \& Kamaya 2007; Inoue \& Inutsuka 2008; Shadmehri et al.
2010). Solar prominences are supposed to be formed as a result of thermal
instability (e.g., Field 1965; Nakagawa, 1970; Heyvaerts 1974; Mason \& Bessey
1983; Karpen et al. 1989). In galaxy clusters, this instability, including the
presence of the magnetic field, was studied in (e.g., Field 1965; Loewenstein
1990; Balbus 1991; Bogdanovi\'{c} et al. 2009; Parrish et al. 2009; Sharma et
al. 2010). The nonlinear stage of thermal instability resulting in formation
of nonlinear cool structures was investigated in the ISM (e.g., Trevisan \&
Ib\'{a}\~{n}ez 2000; S\'{a}nchez-Salcedo et al. 2002; Yatou \& Toh 2009) and
solar corona (Mason \& Bessey 1983; Karpen et al. 1989; Trevisan \&
Ib\'{a}\~{n}ez 2000).

In papers studying thermal instability in astrophysical objects with the
magnetic field, the one-fluid ideal MHD is generally used. The two-fluid model
of the ideal MHD has been treated, e.g., by Fukue \& Kamaya (2007) and Inoue
\& Inutsuka (2008). The non-ideal effects in the magnetic induction equation
have been considered by several authors (e.g., Heyvaerts 1974; Stiele et al.
2006; Shadmehri et al. 2010).

For astrophysical media consisting of many kinds of species (electrons, ions, dust grains, neutrals, and so on), the multicomponent approach considering the dynamics of each species separately is an adequate method of investigation (e.g., Nekrasov 2009a, 2009b, 2009c). The thermal instability in multicomponent media has been
studied by Kopp et al. (1997), Pandey \& Krishan (2001), Pandey et al. (2003),
Shukla \& Sandberg (2003), Kopp \& Shchekinov (2007). Analytical investigation
of thermal instability in multicomponent magnetized media with such
physical effects as collisions between different species, ionization and
recombination, dust charge dynamics, gravity, self-gravity, and so on is a
sufficiently difficult problem. Therefore, one usually treats simplified
models such as, for example, potential perturbations in nonmagnetized (Kopp et
al. 1997; Pandey \& Krishan 2001; Ib\'{a}\~{n}ez \& Shchekinov 2002; Pandey et
al. 2003; Shukla \& Sandberg 2003; Kopp \& Shchekinov 2007) and magnetized
(Kopp et al. 1997; Shukla \& Sandberg 2003) plasmas.

When studying thermal instability, one usually does not take into account
an energy exchange between species in thermal equations. It may be done
at a weak or strong collisional coupling of species. However, an intermediate
case can in general also occur. The inclusion of this effect results in
considerable analytical complications (e.g., Birk 2000; Birk \& Wiechen 2001).
The absence of thermodynamical equilibrium is an additional factor
complicating a problem. However, different temperatures of species can be
observed, for example, in galaxy clusters (Markevitch et al. 1996; Fox \& Loeb
1997; Ettori \& Fabian 1998; Takizawa 1998). Therefore, this effect needs also
to be taken into consideration. When background temperatures of species
are different, it is necessary to take into account the perturbation of the
energy exchange frequency which depends on the number density and temperature.

In this paper, thermal instability in the electron-ion magnetized
plasma relevant to galaxy clusters, solar corona, and other two-component astrophysical objects is investigated. We apply the multicomponent plasma approach when the dynamics of all the species are considered separately through electric field perturbations (the $\mathbf{E}$-approach; see, e.g., Nekrasov 2009a, 2009b,
2009c; Nekrasov \& Shadmehri 2010, 2011). General expressions obtained in
this paper can be applied for a wide range of astrophysical and laboratory
plasmas also containing the neutrals and dust grains. We assume that
background temperatures of electrons and ions are different and include
the energy exchange in thermal equations. We take into account the dependence of  energy exchange collision frequency on density and temperature perturbations. The cooling-heating functions are also considered for both electrons and ions. We do not include ionization and recombination effects and the gravity. Expressions for electron and ion perturbations are obtained in the general form which can be used for other species. As a specific case, we here treat a condensation mode of thermal
instability of perturbations elongated enough along the background magnetic
field. In this case, the dynamical time is smaller than a time during of which
the particles cover the longitudinal wavelength due to their thermal velocity. The opposite, fast sound speed limit, is considered in (Nekrasov 2011).
We derive the general dispersion relation, taking into account the effects
mentioned above, and discuss the limiting cases.

The paper is organized in the following manner. In Section 2, we give
fundamental equations used in this paper. An equilibrium state is considered
in Section 3. General equations for temperature perturbations are obtained in Section 4. In Section 5, equations for components of velocity perturbations are given in the fast dynamical regime. Components of perturbed current are
calculated in Section 6. These components for the simplified collision
contribution are given in Section 7. In Section 8, we derive the dispersion
relation. Its limiting cases are considered in Section 9. We discuss the
obtained results in Section 10. The possible astrophysical implications are
considered in Section 11. Summing up of main points is given in Section 12.

\textbf{\bigskip}

\section{BASIC EQUATIONS}

The fundamental equations we use are
\begin{equation}
\frac{\partial\mathbf{v}_{j}}{\partial t}+\mathbf{v}_{j}\cdot\mathbf{\nabla
v}_{j}=-\frac{\mathbf{\nabla}p_{j}}{m_{j}n_{j}}+\mathbf{F}_{j}\mathbf{+}%
\frac{q_{j}}{m_{j}c}\mathbf{v}_{j}\times\mathbf{B},
\end{equation}
the equation of motion,%
\begin{equation}
\frac{\partial n_{j}}{\partial t}+\mathbf{\nabla}\cdot n_{j}\mathbf{v}_{j}=0,
\end{equation}
the continuity equation,%
\begin{equation}
\frac{\partial T_{i}}{\partial t}+\mathbf{v}_{i}\cdot\mathbf{\nabla}%
T_{i}+\left(  \gamma-1\right)  T_{i}\mathbf{\nabla}\cdot\mathbf{v}%
_{i}=-\left(  \gamma-1\right)  \frac{1}{n_{i}}%
\mathcal{L}%
_{i}\left(  n_{i},T_{i}\right)  +\nu_{ie}^{\varepsilon}\left(  n_{e}%
,T_{e}\right)  \left(  T_{e}-T_{i}\right)
\end{equation}
and%
\begin{equation}
\frac{\partial T_{e}}{\partial t}+\mathbf{v}_{e}\cdot\mathbf{\nabla}%
T_{e}+\left(  \gamma-1\right)  T_{e}\mathbf{\nabla}\cdot\mathbf{v}%
_{e}=-\left(  \gamma-1\right)  \frac{1}{n_{e}}\mathbf{\nabla\cdot q}%
_{e}-\left(  \gamma-1\right)  \frac{1}{n_{e}}%
\mathcal{L}%
_{e}\left(  n_{e},T_{e}\right)  -\nu_{ei}^{\varepsilon}\left(  n_{i}%
,T_{e}\right)  \left(  T_{e}-T_{i}\right)
\end{equation}
are temperature equations for ions and electrons. In Equations (1) and
(2), the index $j=i,e$ denotes ions and electrons, respectively. The value
$\mathbf{F}_{j}$ in Equation (1) is given by%

\begin{align}
\mathbf{F}_{i}  &  =\frac{q_{i}}{m_{i}}\mathbf{E}-\nu_{ie}\left(
\mathbf{v}_{i}-\mathbf{v}_{e}\right)  ,\\
\mathbf{F}_{e}  &  =\frac{q_{e}}{m_{e}}\mathbf{E}-\nu_{ei}\left(
\mathbf{v}_{e}-\mathbf{v}_{i}\right)  .\nonumber
\end{align}
Other notations in Equations (1)-(5) are the following: $q_{j}$ and $m_{j}$
are the charge and mass of species $j=i,e$, $\mathbf{v}_{j}$ is the
hydrodynamic velocity, $n_{j}$ is the number density, $p_{j}=n_{j}T_{j}$ is
the thermal pressure, $T_{j}$ is the temperature in the energy units,
$\nu_{ie}$ ($\nu_{ei}$) is the collision frequency of ions (electrons) with
electrons (ions), $\nu_{ie}^{\varepsilon}(n_{e},T_{e})=2\nu_{ie}$ ($\nu
_{ei}^{\varepsilon}\left(  n_{i},T_{e}\right)  $) is the frequency of the
thermal energy exchange between ions (electrons) and electrons (ions)
(Braginskii 1965), $n_{i}\nu_{ie}^{\varepsilon}\left(  n_{e},T_{e}\right)
=n_{e}\nu_{ei}^{\varepsilon}\left(  n_{i},T_{e}\right)  $, $\gamma$ is the
ratio of specific heats, $\mathbf{E}$\textbf{\ }and $\mathbf{B}$ are the
electric and magnetic fields, and $c$ is the speed of light in vacuum. The
value $\mathbf{q}_{e}$ in Equation (4) is the electron heat flux associated
with the thermal motion in the system of coordinates where the electron gas is
at rest as a whole (Braginskii 1965). As for the latter, we will consider a
weakly collisional plasma when the electron Larmor radius is much smaller than
the electron collisional mean free path. In this case, the electron heat flux
is mainly directed along the magnetic field,
\begin{equation}
\mathbf{q}_{e}=-\chi_{e}\mathbf{b}\left(  \mathbf{b\cdot\nabla}\right)  T_{e},
\end{equation}
where $\chi_{e}$ is the electron thermal conductivity coefficient and
$\mathbf{b=B/}B$ is the unit vector along the magnetic field. In other
respects, a relation between cyclotron and collision frequencies of species
stays arbitrary in general expressions considered below. We only take into
account the electron heat flux (6) because the corresponding ion thermal
conductivity is considerably smaller (Braginskii 1965). We also assume that
the heat flux in equilibrium is absent. The cooling and heating of plasma
species in Equations (3) and (4) are described by function $%
\mathcal{L}%
_{j}(n_{j},T_{j})=n_{j}^{2}\Lambda_{j}\left(  T_{j}\right)  -n_{j}\Gamma_{j}$,
where $\Lambda_{j}$ and $\Gamma_{j}$ are the cooling and heating functions,
respectively. The form of this function differs from the usually used
cooling-heating function $\pounds $, beginning from the classic paper by Field
(1965). Both functions are connected with each other via equality $%
\mathcal{L}%
_{j}\left(  n_{j},T_{j}\right)  =m_{j}n_{j}\pounds _{j}$. Our choice is
analogous to that as in Begelman \& Zweibel (1994), Pandey \& Krishan (2001),
Shukla \& Sandberg (2003), Bogdanovi\'{c} et al. (2009), Parrish et al.
(2009). The function $\Lambda_{j}\left(  T_{j}\right)  $ can be found, for
example, in Tozzi \& Norman (2001).

Electromagnetic equations are Faraday's
\begin{equation}
\mathbf{\nabla\times E=-}\frac{1}{c}\frac{\partial\mathbf{B}}{\partial t}%
\end{equation}
and Ampere`s
\begin{equation}
\mathbf{\nabla\times B=}\frac{4\pi}{c}\mathbf{j}%
\end{equation}
laws, where $\mathbf{j=}\sum_{j}q_{j}n_{j}\mathbf{v}_{j}.$ We consider wave
processes with typical timescales much larger than the time the light spends
to cover the wavelength of perturbations. In this case, one can neglect the
displacement current in Equation (8) that results in quasi-neutrality for both
electromagnetic and purely electrostatic perturbations. The magnetic field
$\mathbf{B}$ includes the background magnetic field $\mathbf{B}_{0}$, the
magnetic field $\mathbf{B}_{0cur}$ of the background electric current (when is
present), and the perturbed magnetic field.

For generality, we assume in the meanwhile that $n_{i}\neq n_{e}$, having in
mind that some expressions obtained below can be applied for multicomponent plasmas.

\bigskip

\section{EQUILIBRIUM STATE}

At first, we will consider an equilibrium state. We assume that the background
flow (average) velocities of species are absent. We do not here involve an
equilibrium inhomogeneity. Then, thermal equations (3) and (4) in equilibrium take the form%
\begin{align}
\left(  \gamma-1\right)  \frac{1}{n_{i0}}%
\mathcal{L}%
_{i}\left(  n_{i0},T_{i0}\right)  -\nu_{ie}^{\varepsilon}(n_{e0}%
,T_{e0})\left(  T_{e0}-T_{i0}\right)   &  =0,\\
\left(  \gamma-1\right)  \frac{1}{n_{e0}}%
\mathcal{L}%
_{e}\left(  n_{e0},T_{e0}\right)  +\nu_{ei}^{\varepsilon}\left(  n_{i0}%
,T_{e0}\right)  \left(  T_{e0}-T_{i0}\right)   &  =0,\nonumber
\end{align}
where the subscript $0$ denotes equilibrium values.

\bigskip

\section{LINEAR\ EQUATIONS\ FOR TEMPERATURE PERTURBATIONS}

We now consider Equations (3) and (4) in the linear approximation. Applying
the operator $\partial/\partial t$ to Equation (3) and using for ions Equation
(2) to exclude the number density perturbation and Equation (9), we find%
\begin{equation}
D_{1i}T_{i1}-D_{2i}T_{e1}=C_{1i}\mathbf{\nabla}\cdot\mathbf{v}_{i1}%
-C_{2i}\mathbf{\nabla}\cdot\mathbf{v}_{e1},
\end{equation}
where and below the subscript $1$ denotes perturbed values. The operators
and notations introduced in Equation (10) are as follows:%
\begin{align}
D_{1i}  &  =\left(  \frac{\partial}{\partial t}+\Omega_{Ti}+\Omega
_{ie}\right)  \frac{\partial}{\partial t},\\
D_{2i}  &  =\left(  \Omega_{Tie}+\Omega_{ie}\right)  \frac{\partial}{\partial
t},\nonumber\\
C_{1i}  &  =T_{i0}\left[  -\left(  \gamma-1\right)  \frac{\partial}{\partial
t}+\Omega_{ni}-\frac{\left(  T_{e0}-T_{i0}\right)  }{T_{i0}}\Omega
_{ie}\right]  ,\nonumber\\
C_{2i}  &  =\Omega_{ie}\left(  T_{e0}-T_{i0}\right)  .\nonumber
\end{align}
Analogously, we obtain for electrons%
\begin{equation}
D_{1e}T_{e1}-D_{2e}T_{i1}=C_{1e}\mathbf{\nabla}\cdot\mathbf{v}_{e1}%
+C_{2e}\mathbf{\nabla}\cdot\mathbf{v}_{i1},
\end{equation}
where%
\begin{align}
D_{1e}  &  =\left(  \frac{\partial}{\partial t}+\Omega_{\chi}+\Omega
_{Te}+\Omega_{Tei}+\Omega_{ei}\right)  \frac{\partial}{\partial t},\\
D_{2e}  &  =\Omega_{ei}\frac{\partial}{\partial t},\nonumber\\
C_{1e}  &  =T_{e0}\left[  -\left(  \gamma-1\right)  \frac{\partial}{\partial
t}+\Omega_{ne}+\frac{\left(  T_{e0}-T_{i0}\right)  }{T_{e0}}\Omega
_{ei}\right]  ,\nonumber\\
C_{2e}  &  =\Omega_{ei}\left(  T_{e0}-T_{i0}\right)  .\nonumber
\end{align}
In notations (11) and (13), we have introduced the following frequencies:%
\begin{align}
\Omega_{\chi}  &  =-\left(  \gamma-1\right)  \frac{\chi_{e0}}{n_{e0}}%
\frac{\partial^{2}}{\partial z^{2}},\\
\Omega_{Te}  &  =\left(  \gamma-1\right)  \frac{\partial%
\mathcal{L}%
_{e}\left(  n_{e0},T_{e0}\right)  }{n_{e0}\partial T_{e0}},\Omega_{Ti}=\left(
\gamma-1\right)  \frac{\partial%
\mathcal{L}%
_{i}\left(  n_{i0},T_{i0}\right)  }{n_{i0}\partial T_{i0}},\nonumber\\
\Omega_{ne}  &  =\left(  \gamma-1\right)  \frac{\partial%
\mathcal{L}%
_{e}\left(  n_{e0},T_{e0}\right)  }{T_{e0}\partial n_{e0}},\Omega_{ni}=\left(
\gamma-1\right)  \frac{\partial%
\mathcal{L}%
_{i}\left(  n_{i0},T_{i0}\right)  }{T_{i0}\partial n_{i0}},\nonumber\\
\Omega_{ei}  &  =\nu_{ei}^{\varepsilon}\left(  n_{i0},T_{e0}\right)
,\Omega_{ie}=\nu_{ie}^{\varepsilon}\left(  n_{e0},T_{e0}\right)  ,\nonumber\\
\Omega_{Tei}  &  =\frac{\partial\nu_{ei}^{\varepsilon}\left(  n_{i0}%
,T_{e0}\right)  }{\partial T_{e0}}\left(  T_{e0}-T_{i0}\right)  ,\Omega
_{Tie}=\frac{\partial\nu_{ie}^{\varepsilon}\left(  n_{e0},T_{e0}\right)
}{\partial T_{e0}}\left(  T_{e0}-T_{i0}\right)  .\nonumber
\end{align}
We assume that the background magnetic field $\mathbf{B}_{0}$ is directed
along the $z$-axis. In notations (11) and (13), we have used the equilibrium
state and the number density dependence of $\nu_{ei}^{\varepsilon}\left(
n_{i0},T_{e0}\right)  \sim n_{i0}$ and $\nu_{ie}^{\varepsilon}\left(
n_{e0},T_{e0}\right)  \sim n_{e0}$. We see from Equations (10) and (12) that
temperature perturbations are connected with a velocity divergence.
Solutions for $T_{e1}$ and $T_{i1}$ are given by%
\begin{equation}
DT_{e1}=G_{1}\ \mathbf{\nabla}\cdot\mathbf{v}_{e1}+G_{2}\mathbf{\nabla}%
\cdot\mathbf{v}_{i1},
\end{equation}%
\begin{equation}
DT_{i1}=G_{3}\mathbf{\nabla}\cdot\mathbf{v}_{e1}+G_{4}\mathbf{\nabla}%
\cdot\mathbf{v}_{i1},
\end{equation}
where the following notations are introduced:%
\begin{align}
D  &  =\left(  D_{1i}D_{1e}-D_{2i}D_{2e}\right)  ,\\
G_{1}  &  =\left(  D_{1i}C_{1e}-D_{2e}C_{2i}\right)  ,\nonumber\\
G_{2}  &  =\left(  D_{1i}C_{2e}+D_{2e}C_{1i}\right)  ,\nonumber\\
G_{3}  &  =\left(  D_{2i}C_{1e}-D_{1e}C_{2i}\right)  ,\nonumber\\
G_{4}  &  =\left(  D_{1e}C_{1i}+D_{2i}C_{2e}\right)  .\nonumber
\end{align}

To find the temperature perturbation $T_{j1}$, we must have expressions for
$\mathbf{\nabla}\cdot\mathbf{v}_{j1}$. General equations for the velocity
$\mathbf{v}_{j1}$ and $\mathbf{\nabla}\cdot\mathbf{v}_{j1}$ are derived in the
Appendix, where expressions for $D$ and $G_{l}$, $l=1,2,3,4$, are also given.
In their general form, the components of $\mathbf{v}_{j1}$ are very complex.
Therefore to proceed further analytically, we here restrict ourselves to a
limiting case in which the dynamical time $\left(  \partial/\partial t\right)  ^{-1}$ is short in comparison with a time the thermal particles need to cover the wavelength along the magnetic field. Some additional simplifying conditions which are satisfied in magnetized plasmas are also used.

We note that the opposite case when the dynamical frequency is smaller then the corresponding sound frequency has been considered in (Nekrasov 2011). The general expressions are the same in the last and this papers. However for convenience of reading, we keep their here.

\bigskip

\section{SPECIFIC CASE: $\frac{\partial^{2}}{\partial t^{2}}\gg\frac
{\partial^{2}}{\partial z^{2}}\left(  v_{Te}^{2}+v_{Ti}^{2}\right)  $}

Equations (A27), (A28) and (A30) are written in their general form, which
allows us to consider different simplified specific cases corresponding to real astrophysical conditions. We further proceed with sufficiently fast
perturbations such that%
\begin{equation}
\frac{\partial^{2}}{\partial t^{2}}\gg\frac{\partial^{2}}{\partial z^{2}%
}\left(  v_{Te}^{2}+v_{Ti}^{2}\right)  .
\end{equation}
This condition is opposite to the one for the fast sound regime (Nekrasov 2011) and
corresponds to the long-wavelength perturbations along the magnetic field. At
the same time, we assume that $\omega_{ci}^{2}\gg\partial^{2}/\partial t^{2}$
for magnetized plasma. Other condition is a common one for hydromagnetic
description, i.e.
\begin{equation}
1\gg\frac{\partial^{2}}{\partial y^{2}}\ \left(  \frac{v_{Te}^{2}}{\omega
_{ce}^{2}}+\frac{v_{Ti}^{2}}{\omega_{ci}^{2}}\right)  ,
\end{equation}
when the Larmor radius of species is much smaller than the transverse
wavelength of perturbations. The square of the velocity $v_{Tj}^{2}$ has the following
estimation:%
\begin{align}
v_{Te}^{2}  &  =\frac{T_{e0}\left(  \frac{\partial}{\partial t}+\Omega
_{ie}\right)  +T_{i0}\Omega_{ei}}{m_{e}\left(  \frac{\partial}{\partial
t}+\Omega_{ie}+\Omega_{ei}\right)  },\\
v_{Ti}^{2}  &  =\frac{T_{i0}\left(  \frac{\partial}{\partial t}+\Omega
_{ei}\right)  +T_{e0}\Omega_{ie}}{m_{i}\left(  \frac{\partial}{\partial
t}+\Omega_{ie}+\Omega_{ei}\right)  }.\nonumber
\end{align}
Expressions (20) are given in the approximate form to unite two cases,
$\frac{\partial}{\partial t}>($or $<)\Omega_{ie,ei}$. We note that operators
$\left(  \partial/\partial t\right)  ^{-1}$ and $\left(  \partial
/\partial\mathbf{r}\right)  ^{-1}$ in expressions (18)-(20) and corresponding
expressions below denote typical dynamical times and wavelengths of
perturbations. Under conditions (18) and (19) and using notations (A10),
(A30), and (A31), we find equations for $P_{i,e1}$(see Equations (A27) and
(A28)):%
\begin{equation}
P_{i1}=\lambda_{i}\left(  -\frac{1}{\omega_{ci}}\frac{\partial^{2}F_{i1x}%
}{\partial y\partial t}+\frac{\partial F_{i1z}}{\partial z}\right)  -\mu
_{i}\left(  -\frac{1}{\omega_{ce}}\frac{\partial^{2}F_{e1x}}{\partial
y\partial t}+\frac{\partial F_{e1z}}{\partial z}\right)  ,
\end{equation}%
\begin{equation}
P_{e1}=\lambda_{e}\left(  -\frac{1}{\omega_{ce}}\frac{\partial^{2}F_{e1x}%
}{\partial y\partial t}+\frac{\partial F_{e1z}}{\partial z}\right)  -\mu
_{e}\left(  -\frac{1}{\omega_{ci}}\frac{\partial^{2}F_{i1x}}{\partial
y\partial t}+\frac{\partial F_{i1z}}{\partial z}\right)  ,
\end{equation}
where the following notations are introduced:%
\begin{align}
\lambda_{i}  &  =\frac{T_{i0}}{m_{i}}\left(  \frac{\partial}{\partial
t}\right)  ^{-1}-\frac{G_{4}}{Dm_{i}},\mu_{i}=\frac{G_{3}}{Dm_{i}},\\
\lambda_{e}  &  =\frac{T_{e0}}{m_{e}}\left(  \frac{\partial}{\partial
t}\right)  ^{-1}-\frac{G_{1}}{Dm_{e}},\mu_{e}=\frac{G_{2}}{Dm_{e}}.\nonumber
\end{align}
Equations (21) and (22) have a symmetric form relatively to changing the index
$i$ by $e$ and vice versa. Estimations of $\lambda_{j}$ and $\mu_{j}$ are
followed from expressions (A16), (A18), (A19), and (A29).

\subsection{EQUATIONS FOR COMPONENTS OF VELOCITIES $\mathbf{v}_{i,e1}$}

We now obtain equations for components of velocities $\mathbf{v}_{i,e1}$,
using Equations (21) and (22).

\subsubsection{Equations for $v_{i,e1y}$}

From Equations (A3), (21), and (22), we find, using notations (A6),%
\begin{align}
v_{i1y}  &  =-\frac{1}{\omega_{ci}}F_{i1x}+\frac{1}{\omega_{ci}^{3}}%
\frac{\partial^{2}F_{i1x}}{\partial t^{2}}+\frac{1}{\omega_{ci}^{2}}%
\frac{\partial F_{i1y}}{\partial t}-\frac{1}{\omega_{ci}^{4}}\frac
{\partial^{3}F_{i1y}}{\partial t^{3}}\\
&  -\frac{1}{\omega_{ci}^{2}}\frac{\partial^{3}}{\partial y^{2}\partial
t}\left(  \frac{\lambda_{i}}{\omega_{ci}}F_{i1x}-\frac{\mu_{i}}{\omega_{ce}%
}F_{e1x}\right) \nonumber\\
&  +\frac{1}{\omega_{ci}^{2}}\frac{\partial^{2}}{\partial y\partial z}\left(
\lambda_{i}F_{i1z}-\mu_{i}F_{e1z}\right)  ,\nonumber
\end{align}%
\begin{align}
v_{e1y}  &  =-\frac{1}{\omega_{ce}}F_{e1x}+\frac{1}{\omega_{ce}^{3}}%
\frac{\partial^{2}F_{e1x}}{\partial t^{2}}+\frac{1}{\omega_{ce}^{2}}%
\frac{\partial F_{e1y}}{\partial t}-\frac{1}{\omega_{ce}^{4}}\frac
{\partial^{3}F_{e1y}}{\partial t^{3}}\\
&  -\frac{1}{\omega_{ce}^{2}}\frac{\partial^{3}}{\partial y^{2}\partial
t}\left(  \frac{\lambda_{e}}{\omega_{ce}}F_{e1x}-\frac{\mu_{e}}{\omega_{ci}%
}F_{i1x}\right) \nonumber\\
&  +\frac{1}{\omega_{ce}^{2}}\frac{\partial^{2}}{\partial y\partial z}\left(
\lambda_{e}F_{e1z}-\mu_{e}F_{i1z}\right)  .\nonumber
\end{align}
The terms proportional to $\omega_{cj}^{-4}$ are needed in equations for
$v_{i,e1x}$ for obtaining terms $\sim\omega_{cj}^{-3}$. We see that these
solutions are obtained one from another by changing $i\leftrightarrow e$.

\subsubsection{Equations for $v_{i,e1x}$}

Equations for $v_{i,e1x}$ are easily found from Equation (A2) by using
Equations (24) and (25):%
\begin{align}
v_{i1x}  &  =\frac{1}{\omega_{ci}}F_{i1y}+\frac{1}{\omega_{ci}^{2}}%
\frac{\partial F_{i1x}}{\partial t}-\frac{1}{\omega_{ci}^{3}}\frac
{\partial^{2}F_{i1y}}{\partial t^{2}}\\
&  -\frac{1}{\omega_{ci}}\frac{\partial^{2}}{\partial y^{2}}\left(
\frac{\lambda_{i}}{\omega_{ci}}F_{i1x}-\frac{\mu_{i}}{\omega_{ce}}%
F_{e1x}\right) \nonumber\\
&  +\frac{1}{\omega_{ci}}\frac{\partial^{2}}{\partial y\partial z}\left(
\frac{\partial}{\partial t}\right)  ^{-1}\left(  \lambda_{i}F_{i1z}-\mu
_{i}F_{e1z}\right)  ,\nonumber
\end{align}%
\begin{align}
v_{e1x}  &  =\frac{1}{\omega_{ce}}F_{e1y}+\frac{1}{\omega_{ce}^{2}}%
\frac{\partial F_{e1x}}{\partial t}-\frac{1}{\omega_{ce}^{3}}\frac
{\partial^{2}F_{e1y}}{\partial t^{2}}\\
&  -\frac{1}{\omega_{ce}}\frac{\partial^{2}}{\partial y^{2}}\left(
\frac{\lambda_{e}}{\omega_{ce}}F_{e1x}-\frac{\mu_{e}}{\omega_{ci}}%
F_{i1x}\right) \nonumber\\
&  +\frac{1}{\omega_{ce}}\frac{\partial^{2}}{\partial y\partial z}\left(
\frac{\partial}{\partial t}\right)  ^{-1}\left(  \lambda_{e}F_{e1z}-\mu
_{e}F_{i1z}\right)  .\nonumber
\end{align}

\subsubsection{Equations for $v_{i,e1z}$}

From Equations (A7), (21), and (22), we obtain, using notations (A6),%
\begin{align}
\frac{\partial^{2}v_{i1z}}{\partial t^{2}}  &  =\frac{\partial F_{i1z}%
}{\partial t}-\frac{\partial^{3}}{\partial y\partial z\partial t}\left(
\frac{\lambda_{i}}{\omega_{ci}}F_{i1x}-\frac{\mu_{i}}{\omega_{ce}}%
F_{e1x}\right) \\
&  +\frac{\partial^{2}}{\partial z^{2}}\left(  \lambda_{i}F_{i1z}-\mu
_{i}F_{e1z}\right)  ,\nonumber
\end{align}%
\begin{align}
\frac{\partial^{2}v_{e1z}}{\partial t^{2}}  &  =\frac{\partial F_{e1z}%
}{\partial t}-\frac{\partial^{3}}{\partial y\partial z\partial t}\left(
\frac{\lambda_{e}}{\omega_{ce}}F_{e1x}-\frac{\mu_{e}}{\omega_{ci}}%
F_{i1x}\right) \\
&  +\frac{\partial^{2}}{\partial z^{2}}\left(  \lambda_{e}F_{e1z}-\mu
_{e}F_{i1z}\right)  .\nonumber
\end{align}

\section{COMPONENTS OF CURRENT}

We now find components of the linear current $\mathbf{j}_{1}=%
{\displaystyle\sum\limits_{j}}
q_{j}n_{j0}\mathbf{v}_{j1}$. It is convenient to consider the value
$4\pi\left(  \partial/\partial t\right)  ^{-1}\mathbf{j}_{1}$. In this case,
we obtain dimensionless coefficients by $\mathbf{E}_{1}$. We further consider
the electron-ion plasma in which $n_{e0}=n_{i0}=n_{0}$, $q_{e}=-q_{i}$. In our
calculations, we will use an equality $m_{e}\nu_{ei}=m_{i}\nu_{ie}$ and the
relation $m_{i}\mathbf{F}_{i1}=-m_{e}\mathbf{F}_{e1}$. From Equations
(24)-(29) and using notations (5) in the linear approximation, we find%
\begin{align}
4\pi\left(  \frac{\partial}{\partial t}\right)  ^{-1}j_{1x}  &  =a_{xx}%
E_{1x}-a_{xy}E_{1y}+a_{xz}E_{1z}\\
&  -b_{xx}\left(  v_{i1x}-v_{e1x}\right)  +b_{xy}\left(  v_{i1y}%
-v_{e1y}\right)  -b_{xz}\left(  v_{i1z}-v_{e1z}\right)  ,\nonumber
\end{align}%
\begin{align}
4\pi\left(  \frac{\partial}{\partial t}\right)  ^{-1}j_{1y}  &  =a_{yx}%
E_{1x}+a_{yy}E_{1y}+a_{yz}E_{1z}\\
&  -b_{yx}\left(  v_{i1x}-v_{e1x}\right)  -b_{yy}\left(  v_{i1y}%
-v_{e1y}\right)  -b_{yz}\left(  v_{i1z}-v_{e1z}\right)  ,\nonumber
\end{align}%
\begin{align}
4\pi\left(  \frac{\partial}{\partial t}\right)  ^{-1}j_{1z}  &  =a_{zx}%
E_{1x}+a_{zz}E_{1z}\\
&  -b_{zx}\left(  v_{i1x}-v_{e1x}\right)  -b_{zz}\left(  v_{i1z}%
-v_{e1z}\right)  ,\nonumber
\end{align}
Here, the following notations are introduced:%
\begin{align}
a_{xx}  &  =\frac{\omega_{pi}^{2}}{\omega_{ci}^{2}}\left\{  1-\frac{1}{m_{i}%
}\left[  \left(  \lambda_{e}-\mu_{e}\right)  m_{e}+\left(  \lambda_{i}-\mu
_{i}\right)  m_{i}\right]  \frac{\partial^{2}}{\partial y^{2}}\left(
\frac{\partial}{\partial t}\right)  ^{-1}\right\}  ,\\
a_{xy}  &  =\frac{\omega_{pi}^{2}}{\omega_{ci}^{3}}\frac{\partial}{\partial
t},a_{xz}=\frac{\omega_{pi}^{2}}{\omega_{ci}}\left[  \frac{\lambda_{i}%
m_{i}-\mu_{e}m_{e}}{m_{i}}-\frac{\lambda_{e}m_{e}-\mu_{i}m_{i}}{m_{e}}\right]
\frac{\partial^{2}}{\partial y\partial z}\left(  \frac{\partial}{\partial
t}\right)  ^{-2},\nonumber\\
a_{yx}  &  =\frac{\omega_{pi}^{2}}{\omega_{ci}^{2}}\left\{  \frac{1}%
{\omega_{ci}}\frac{\partial}{\partial t}-\frac{1}{m_{i}}\left[  \frac{\left(
\lambda_{i}-\mu_{i}\right)  m_{i}}{\omega_{ci}}+\frac{\left(  \lambda_{e}%
-\mu_{e}\right)  m_{e}}{\omega_{ce}}\right]  \frac{\partial^{2}}{\partial
y^{2}}\right\}  ,a_{yy}=\frac{\omega_{pi}^{2}}{\omega_{ci}^{2}},\nonumber\\
a_{yz}  &  =\frac{\omega_{pi}^{2}}{\omega_{ci}}\left[  \frac{1}{\omega_{ci}%
}\left(  \lambda_{i}+\mu_{i}\frac{m_{i}}{m_{e}}\right)  -\frac{1}{\omega_{ce}%
}\left(  \lambda_{e}+\mu_{e}\frac{m_{e}}{m_{i}}\right)  \right]
\frac{\partial^{2}}{\partial y\partial z}\left(  \frac{\partial}{\partial
t}\right)  ^{-1},\nonumber\\
a_{zx}  &  =\frac{\omega_{pi}^{2}}{\omega_{ci}}\left[  \left(  \lambda_{e}%
-\mu_{e}\right)  -\left(  \lambda_{i}-\mu_{i}\right)  \right]  \frac
{\partial^{2}}{\partial y\partial z}\left(  \frac{\partial}{\partial
t}\right)  ^{-2},a_{zy}=0,a_{zz}=\omega_{pi}^{2}\frac{m_{i}}{m_{e}}\left(
\frac{\partial}{\partial t}\right)  ^{-2},\nonumber\\
b_{ij}  &  =a_{ij}\frac{m_{i}}{q_{i}}\nu_{ie}.\nonumber
\end{align}
where $\omega_{pi}=\left(  4\pi n_{i0}q_{i}^{2}/m_{i}\right)  ^{1/2}$ is the
ion plasma frequency.

\bigskip

\section{SIMPLIFICATION\ OF\ COLLISION\ CONTRIBUTION}

Relationship between $\omega_{ce}$ and $\nu_{ei}$ or $\omega_{ci}$ and
$\nu_{ie}$ (that is the same) can be arbitrary in Equations (30)-(32) (except
of that in the thermal conduction). We further proceed by taking into account
that $\partial/\partial t\ll\omega_{ci}$. In this case, we can neglect collisional terms proportional to $b_{xy}$ and $b_{yx}$ (see notations (33)).
However, a system of Equations (30)-(32) stays sufficiently complex to find
$\mathbf{j}_{1}$ through $\mathbf{E}_{1}$. Therefore, we further consider the
case in which the following condition is satisfied:%
\begin{equation}
1\gg\frac{\nu_{ie}}{\omega_{ci}^{2}}\frac{\partial}{\partial t}\left\{
1-\frac{1}{m_{i}}\left[  m_{e}\left(  \lambda_{e}-\mu_{e}\right)
+m_{i}\left(  \lambda_{i}-\mu_{i}\right)  \right]  \frac{\partial^{2}%
}{\partial y^{2}}\left(  \frac{\partial}{\partial t}\right)  ^{-1}\right\}
\end{equation}
or on the order of magnitude
\[
1\gg\frac{\nu_{ie}}{\omega_{ci}^{2}}\frac{\partial}{\partial t}\left[
1+c_{s}^{2}\frac{\partial^{2}}{\partial y^{2}}\left(  \frac{\partial}{\partial
t}\right)  ^{-2}\right]  ,
\]
where $c_{s}^{2}=\left(  T_{e0}+T_{i0}\right)  /m_{i}$. It is obvious that
this inequality can easily be realized in magnetized plasma. Under condition
(34), we can neglect the terms $\sim b_{xx}$ and $b_{zx}$ in Equations (30)
and (32). In the case $\omega_{ci}^{2}\gg\nu_{ie}\partial/\partial t$, the
term $\sim b_{yy}$ in Equation (31) can also be omitted. Thus, a system of
Equations (30)-(32) takes the form,%
\begin{equation}
4\pi\left(  \frac{\partial}{\partial t}\right)  ^{-1}j_{1x}=\varepsilon
_{xx}E_{1x}-\varepsilon_{xy}E_{1y}+\varepsilon_{xz}E_{1z},
\end{equation}%
\begin{equation}
4\pi\left(  \frac{\partial}{\partial t}\right)  ^{-1}j_{1y}=\varepsilon
_{yx}E_{1x}+\varepsilon_{yy}E_{1y}+\varepsilon_{yz}E_{1z},
\end{equation}%
\begin{equation}
4\pi\left(  \frac{\partial}{\partial t}\right)  ^{-1}j_{1z}=\varepsilon
_{zx}E_{1x}+\varepsilon_{zz}E_{1z}.
\end{equation}
The following notations are here introduced:%
\begin{align}
\varepsilon_{xx}  &  =a_{xx}-\frac{\nu_{ie}}{\omega_{pi}^{2}}\frac{\partial
}{\partial t}\frac{a_{xz}a_{zx}}{\left(  1+d\right)  },\varepsilon_{xy}%
=a_{xy},\varepsilon_{xz}=\frac{a_{xz}}{\left(  1+d\right)  },\\
\varepsilon_{yx}  &  =a_{yx}-\frac{\nu_{ie}}{\omega_{pi}^{2}}\frac{\partial
}{\partial t}\frac{a_{yz}a_{zx}}{\left(  1+d\right)  },\varepsilon_{yy}%
=a_{yy},\varepsilon_{yz}=\frac{a_{yz}}{\left(  1+d\right)  },\nonumber\\
\varepsilon_{zx}  &  =\frac{a_{zx}}{\left(  1+d\right)  },\varepsilon
_{zz}=\frac{a_{zz}}{\left(  1+d\right)  },\nonumber
\end{align}
where%
\begin{equation}
d=a_{zz}\frac{\nu_{ie}}{\omega_{pi}^{2}}\frac{\partial}{\partial t}=\nu
_{ei}\left(  \frac{\partial}{\partial t}\right)  ^{-1}.
\end{equation}
Parameter $d$ defines the collisionless, $d\ll1$, and collisional, $d\gg1$,
regimes. Below, we derive the dispersion relation.

\bigskip

\section{DISPERSION\ RELATION AND\ ELECTRIC\ FIELD POLARIZATION\ }

We further consider Equations (35)-(37) in the Fourier-representation,
assuming that perturbations have the form $\exp\left(  i\mathbf{k\cdot
r-}i\omega t\right)  $. Then using Equations (7) and (8), we obtain the
following system of equations:%
\begin{align}
\left(  n^{2}-\varepsilon_{xx}\right)  E_{1xk}+\varepsilon_{xy}E_{1yk}%
-\varepsilon_{xz}E_{1zk}  &  =0,\\
-\varepsilon_{yx}E_{1xk}+\left(  n_{z}^{2}-\varepsilon_{yy}\right)
E_{1yk}-\left(  n_{y}n_{z}+\varepsilon_{yz}\right)  E_{1zk}  &  =0,\nonumber\\
-\varepsilon_{zx}E_{1xk}-n_{y}n_{z}E_{1yk}+\left(  n_{y}^{2}-\varepsilon
_{zz}\right)  E_{1zk}  &  =0,\nonumber
\end{align}
where $\mathbf{E}_{1k}$ is the Fourier-image of the electric field
perturbation, $\mathbf{n=k}c/\omega$. The index $k$ by $\mathbf{E}_{1k}$ is
equal to $k=\left\{  \mathbf{k,}\omega\right\}  $. For the Fourier-images of
operators $\varepsilon_{ij}$ and $d$, we keep the same notations. In general,
we see that the longitudinal electric field $E_{1zk}\sim E_{1x,yk}$ inevitably
arises when $k_{y}\neq0$ and $n_{y}^{2}-\varepsilon_{zz}\neq0$. The dispersion
relation can be found by setting the determinant of the system (40) equal to zero.

We will consider the case in which
\begin{equation}
\varepsilon_{zz}\gg n_{y}^{2}.
\end{equation}
In the collisionless regime ($d\ll1$), this inequality denotes that the
transverse wavelength of perturbations is much larger than the electron
skin-depth. We assume that condition (41) is also satisfied in the collisional
regime ($d\gg1$). Expressing the electric field $E_{1zk}$ through $E_{1x,yk}$
from the third Equation (40) and substituting it into two other equations, we
find, using expressions (33) and (38), that contribution of the longitudinal
electric field $E_{1zk}$ in the case (18) is negligible. For an estimation of
values (33), we take%
\begin{equation}
\left(  \lambda_{j}-\mu_{j}\right)  m_{j}\sim\left(  T_{e0}+T_{i0}\right)
\left(  \frac{\partial}{\partial t}\right)  ^{-1}%
\end{equation}
(see expressions (23)). From Equations (33) and (38), it is easy to see that
$\varepsilon_{yx}/\varepsilon_{xx}\sim\omega/\omega_{ci}$ and $\varepsilon
_{xy}/\varepsilon_{yy}\sim\omega/\omega_{ci}$. Since $\omega^{2}\ll\omega
_{ci}^{2}$, the contribution of term $\varepsilon_{xy}\varepsilon_{yx}$ into
dispersion relation is small as compared with that of term $\varepsilon
_{xx}\varepsilon_{yy}$. Then using an estimation (42), we obtain\ the simple
dispersion relation
\begin{equation}
\left(  n_{z}^{2}-\varepsilon_{yy}\right)  \left(  n^{2}-\varepsilon
_{xx}\right)  =0.
\end{equation}
The first factor of Equation (43) describes the Alfv\'{e}n wave $\omega
^{2}=k_{z}^{2}c_{A}^{2}$, where $c_{A}=B_{0}/\left(  4\pi m_{i}n_{i0}\right)
^{1/2}$ is the Alfv\'{e}n velocity. The polarization of this wave is
$\mathbf{E}_{1}=\left(  0,E_{1y},0\right)  $. The second factor of Equation
(43),%
\begin{equation}
n^{2}-\varepsilon_{xx}=0,
\end{equation}
describes the magnetosonic kind of perturbations with polarization
$\mathbf{E}_{1}=\left(  E_{1x},0,0\right)  $. This perturbation is purely the
electromagnetic one.

\bigskip

\section{\bigskip SOLUTION OF DISPERSION\ RELATION (44)}

The parameter $d$ defined by Equation (39) is equal to $d=i\nu_{ei}/\omega$.
If $d\ll1$, then the collisional term in $\varepsilon_{xx}$ is imaginary one
and of the order of $\left(  \nu_{ei}/\omega\right)  \left(  k_{z}^{2}%
v_{Te}^{2}/\omega^{2}\right)  \ll1$ in comparison with the term $a_{xx}$. In
the case $d\gg1$, this collisional term becomes the real one and is $\sim$
$\left(  k_{z}^{2}v_{Te}^{2}/\omega^{2}\right)  \ll1$ in comparison with
$a_{xx}$. Therefore, the collisional term in $\varepsilon_{xx}$ is not taken
into consideration in Equation (44). Then, the dispersion relation takes
the form,%
\begin{equation}
\omega^{2}=k^{2}c_{A}^{2}+k_{y}^{2}\frac{1}{m_{i}D}\left[  D\left(
T_{e0}+T_{i0}\right)  +i\omega\left(  G_{1}+G_{2}+G_{3}+G_{4}\right)  \right]
.
\end{equation}
The values $D$ and $G_{i}$, $i=1,2,3,4$, are given by expressions (A16)-(A20),
where $\partial/\partial t$ should be replaced by $-i\omega$. Calculating
expression in the square brackets in Equation (45), we obtain%
\begin{equation}
\omega^{2}=k^{2}c_{A}^{2}+k_{y}^{2}\frac{R}{m_{i}V}.
\end{equation}
Here,%
\begin{align}
R  &  =T_{e0}\left(  -i\gamma\omega+\Omega_{\chi}+\Omega_{Te}-\Omega
_{ne}\right)  \left(  -i\omega+\Omega_{Ti}+2\Omega_{ie}\right) \\
&  +T_{i0}\left(  -i\gamma\omega+\Omega_{Ti}-\Omega_{ni}\right)  \left(
-i\omega+\Omega_{\chi}+\Omega_{Te}+2\Omega_{ei}\right) \nonumber\\
&  +\left(  T_{e0}-T_{i0}\right)  \left(  -i\omega+\Omega_{\chi}+\Omega
_{Te}\right)  \Omega_{ie}+\left(  T_{i0}-T_{e0}\right)  \left(  -i\omega
+\Omega_{Ti}\right)  \Omega_{ei}\nonumber\\
&  +T_{e0}\left(  -i\omega+\Omega_{Ti}\right)  \Omega_{Tei}+T_{e0}\left[
-i\left(  \gamma-1\right)  \omega-\Omega_{ne}\right]  \Omega_{Tie}%
+T_{i0}\left(  -i\gamma\omega+\Omega_{Ti}-\Omega_{ni}\right)  \Omega
_{Tei}\nonumber
\end{align}
and%
\begin{equation}
V=\left[  \left(  -i\omega+\Omega_{\chi}+\Omega_{Te}\right)  \left(
-i\omega+\Omega_{Ti}+\Omega_{ie}\right)  +\left(  -i\omega+\Omega_{Ti}\right)
\left(  \Omega_{ei}+\Omega_{Tei}\right)  \right]  ,
\end{equation}
where $\Omega_{\chi}=\left(  \gamma-1\right)  \left(  \chi_{e0}/n_{e0}\right)
k_{z}^{2}$. All the values $\Omega$ are defined by a system (14). We see that
the first four terms on the right hand-side of expression (47) are symmetric
ones relatively to contribution of electrons and ions. The last three
terms are connected with perturbation of collision frequency $\nu
_{ei,ie}^{\varepsilon}\left(  n_{i,e0},T_{e}\right)  $ because of the electron
temperature perturbation. All the terms proportional to $\Omega_{ei,ie}$ and
$\Omega_{Tei,ie}$ in expressions (47) and (48) are connected with the energy
exchange in thermal equations (3) and (4). We see that this effect in a
general case when $T_{e0}\neq T_{i0}$ results in considerable modification
(complication) of dispersion relation. Nevertheless, it must be taken into
account because the absence of thermal equilibrium between electrons and
ions can be observed, for example, in the outer part of galaxy clusters (e.g., Markevitch et al. 1996; Fox \& Loeb 1997; Ettori \& Fabian 1998; Takizawa 1998).

Dispersion relation (46) has a general form which permits us to investigate
analytically different limiting cases. When the thermal pressure is larger
than the magnetic pressure, the case which is satisfied in the intracluster
medium, one can omit the first term on the right hand-side of Equation (46).
Then, we can write this equation in the form
\begin{equation}
\frac{\omega^{2}}{k_{y}^{2}c_{s}^{2}}=\frac{R}{\left(  T_{e0}+T_{i0}\right)
V},
\end{equation}
where $c_{s}$ is the sound speed. It is followed from Equation (49) that in
the case $\omega^{2}\gg k_{y}^{2}c_{s}^{2}$ or $R\gg\left(  T_{e0}%
+T_{i0}\right)  V$, dispersion relation is given by%
\begin{equation}
V=0.
\end{equation}
In the opposite case, $\omega^{2}\ll k_{y}^{2}c_{s}^{2}$ or $R\ll\left(
T_{e0}+T_{i0}\right)  V$, we have%
\begin{equation}
R=0.
\end{equation}

\bigskip

\subsection{SOLUTION\ OF\ DISPERSION\ RELATION\ $V=0$}

When $\Omega_{ie,ei}=0$, i.e. at the absence of energy exchange, Equation (50)
gives
\begin{equation}
\left(  -i\omega+\Omega_{\chi}+\Omega_{Te}\right)  \left(  -i\omega
+\Omega_{Ti}\right)  =0.
\end{equation}
Thus, an instability can be generated by the electrons or ions. We see that
Equation (52) has solutions which correspond to the isochoric ones in the MHD  (Parker 1953; Field 1965). When
$\Omega_{ie,ei}\rightarrow\infty$, the case of a strong energy coupling, we
obtain%
\begin{equation}
\omega=-i\frac{\left(  \Omega_{\chi}+\Omega_{Te}\right)  \Omega_{ie}%
+\Omega_{Ti}\left(  \Omega_{ei}+\Omega_{Tei}\right)  }{\left(  \Omega
_{ie}+\Omega_{ei}+\Omega_{Tei}\right)  }.
\end{equation}
Taking into account that $\Omega_{ie}=\Omega_{ei}$ at $n_{e0}=n_{i0}$ and
$\Omega_{Tei}=-3\Omega_{ei}\left(  T_{e0}-T_{i0}\right)  /2T_{e0}$ (Braginskii
1965), this equation becomes the following:
\begin{equation}
\omega=-i\frac{2\Omega_{\chi}+2\Omega_{Te}+\Omega_{Ti}\left(  3T_{i0}%
/T_{e0}-1\right)  }{1+3T_{i0}/T_{e0}}.
\end{equation}
We see that if $3T_{i0}/T_{e0}\ll1$, then $\omega=-i\left(  2\Omega_{\chi
}+2\Omega_{Te}-\Omega_{Ti}\right)  $. Thus, the ions can contribute to
instability when $\Omega_{Ti}>0$. In this case, condition of instability
takes the form%
\[
-\frac{\partial\Lambda_{e}\left(  T_{e0}\right)  }{\partial T_{e0}}%
+\frac{\partial\Lambda_{i}\left(  T_{i0}\right)  }{2\partial T_{i0}}%
>\frac{\chi_{e0}}{n_{0}^{2}}k_{z}^{2},
\]
where we have only taken into account cooling functions of electrons and ions
(see Section 2). We see from this condition that instability can also be possible in the electron temperature
domain where $\partial\Lambda_{e}\left(  T_{e0}\right)  /\partial T_{e0}>0$.

\bigskip

\subsection{SOLUTION\ OF\ DISPERSION\ RELATION\ $R=0$}

We now consider the dispersion relation (51) which is appropriate in the case
$\omega^{2}\ll k_{y}^{2}c_{s}^{2}$. This equation coincides with the dispersion relation in the fast sound speed regime (Nekrasov 2011). For reading convenience, we repeat here results given in the last paper. By using the temperature
dependence of $\nu_{ei,ie}^{\varepsilon}\sim T_{e}^{-3/2}$, Equation (51) can be rewritten in
the form,
\begin{align}
&  T_{e0}\left(  -i\gamma\omega+\Omega_{\chi}+\Omega_{Te}-\Omega_{ne}\right)
\left(  -i\omega+\Omega_{Ti}+2\Omega_{ie}\right) \\
&  +T_{i0}\left(  -i\gamma\omega+\Omega_{Ti}-\Omega_{ni}\right)  \left(
-i\omega+\Omega_{\chi}+\Omega_{Te}+2\Omega_{ie}\right) \nonumber\\
&  +\Omega_{ie}\left(  T_{e0}-T_{i0}\right)  \left(  \Omega_{\chi}+\Omega
_{Te}-\Omega_{Ti}\right) \nonumber\\
&  -\frac{3}{2}\Omega_{ie}\left(  T_{e0}-T_{i0}\right)  \left[  \left(
-i\gamma\omega+\Omega_{Ti}\right)  \left(  1+\frac{T_{i0}}{T_{e0}}\right)
-\Omega_{ne}-\frac{T_{i0}}{T_{e0}}\Omega_{ni}\right]  =0.\nonumber
\end{align}
The different limiting cases of Equation (55) are given in
subsections of 9.2.

\subsubsection{The case $\Omega_{ie}=0$}

If we do not take into account the energy exchange, $\Omega_{ie}=0$, and set
$T_{e0}=T_{i0}$, then we obtain equation%

\begin{align}
&  2\gamma\omega^{2}+i\left[  \left(  \gamma+1\right)  \left(  \Omega_{\chi
}+\Omega_{Te}+\Omega_{Ti}\right)  -\Omega_{ne}-\Omega_{ni}\right]  \omega\\
&  -2\left(  \Omega_{\chi}+\Omega_{Te}\right)  \Omega_{Ti}+\Omega_{ne}%
\Omega_{Ti}+\Omega_{ni}\left(  \Omega_{\chi}+\Omega_{Te}\right)  =0.\nonumber
\end{align}
Neglecting the contribution of the ion cooling and heating, $\Omega
_{Ti}=\Omega_{ni}=0$, we have%
\[
\omega=-\frac{i}{2\gamma}\left[  \left(  \gamma+1\right)  \left(  \Omega
_{\chi}+\Omega_{Te}\right)  -\Omega_{ne}\right]  .
\]
It is easy to see that this solution is a mixture of isochoric and isobaric
solutions (Parker 1953; Field 1965) because we have taken into account the ion
temperature perturbation. If we neglect the latter, i.e. neglect the second
term $\sim T_{i0}$ in Equation (55), we obtain the usual isobaric solution
\[
\omega=-\frac{i}{\gamma}\left(  \Omega_{\chi}+\Omega_{Te}-\Omega_{ne}\right)
.
\]
We also see from Equation (55) that for short-wavelength perturbations
when $\Omega_{\chi}\gg\omega,\Omega_{Te},\Omega_{ne}$ the thermal instability
can arise due to the ion cooling function%
\[
\omega=-\frac{i}{T_{e0}+\gamma T_{i0}}\left[  \left(  T_{e0}+T_{i0}\right)
\Omega_{Ti}-T_{i0}\Omega_{ni}\right]  .
\]

\subsubsection{The case $\Omega_{ie}\rightarrow\infty$}

When the frequency $\Omega_{ie}$ is much larger than other frequencies,
$2\Omega_{ie}\gg\omega,\Omega_{\chi},\Omega_{Te,i}$, and $T_{e0}=T_{i0}$, then
the dispersion relation becomes the following:%
\begin{equation}
\omega=-\frac{i}{2\gamma}\left(  \Omega_{\chi}+\Omega_{Te}-\Omega_{ne}%
+\Omega_{Ti}-\Omega_{ni}\right)  .
\end{equation}
This is isobaric solution with the electron and ion cooling-heating.

For different temperatures of electrons and ions, $T_{e0}\neq T_{i0}$, we obtain%

\begin{align}
i\frac{\gamma}{2}\left[  T_{e0}+\left(  4+3\frac{T_{i0}}{T_{e0}}\right)
T_{i0}\right]  \omega &  =\left(  3T_{e0}-T_{i0}\right)  \left(  \Omega_{\chi
}+\Omega_{Te}\right)  -\left[  \frac{5}{2}T_{e0}-3T_{i0}\left(  1+\frac
{T_{i0}}{2T_{e0}}\right)  \right]  \Omega_{Ti}\\
&  -\frac{1}{2}\left(  3T_{i0}+T_{e0}\right)  \left(  \Omega_{ne}+\frac
{T_{i0}}{T_{e0}}\Omega_{ni}\right)  .\nonumber
\end{align}
In the case $T_{e0}\gg T_{i0}$, this equation takes the form,%
\[
\omega=-\frac{i}{\gamma}\left[  6\left(  \Omega_{\chi}+\Omega_{Te}\right)
-\Omega_{ne}-5\Omega_{Ti}\right]  .
\]
In the opposite case, $T_{e0}\ll T_{i0}$, we obtain the ion isobaric solution%
\[
\omega=-\frac{i}{\gamma}\left(  \Omega_{Ti}-\Omega_{ni}\right)  .
\]

\subsubsection{General case}

In a general case, Equation (55) can be written in the form%
\begin{equation}
g_{0}\omega^{2}+ig_{1}\omega-g_{2}=0,
\end{equation}
where
\begin{align}
g_{0}  &  =\gamma\left(  T_{e0}+T_{i0}\right)  ,\\
g_{1}  &  =\left[  \left(  \gamma T_{i0}+T_{e0}\right)  \left(  \Omega_{\chi
}+\Omega_{Te}\right)  +\left(  \gamma T_{e0}+T_{i0}\right)  \Omega_{Ti}%
-T_{e0}\Omega_{ne}-T_{i0}\Omega_{ni}\right] \nonumber\\
&  +\frac{1}{2}\gamma\left[  T_{e0}+T_{i0}\left(  4+3\frac{T_{i0}}{T_{e0}%
}\right)  \right]  \Omega_{ie},\nonumber\\
g_{2}  &  =T_{e0}\left(  \Omega_{\chi}+\Omega_{Te}-\Omega_{ne}\right)
\Omega_{Ti}+T_{i0}\left(  \Omega_{Ti}-\Omega_{ni}\right)  \left(  \Omega
_{\chi}+\Omega_{Te}\right) \nonumber\\
&  +\left(  3T_{e0}-T_{i0}\right)  \Omega_{ie}\left(  \Omega_{\chi}%
+\Omega_{Te}\right)  -\left[  \frac{5}{2}T_{e0}-3T_{i0}\left(  1+\frac{T_{i0}%
}{2T_{e0}}\right)  \right]  \Omega_{ie}\Omega_{Ti}\nonumber\\
&  -\frac{1}{2}\left(  T_{e0}+3T_{i0}\right)  \Omega_{ie}\left(  \Omega
_{ne}+\Omega_{ni}\frac{T_{i0}}{T_{e0}}\right)  .\nonumber
\end{align}
Equation (59) can be solved numerically for known cooling-heating functions
and temperatures.

\bigskip

\section{DISCUSSION}

Applying the operator $\mathbf{\nabla}\cdot$ to Equation (8), we can see that
$n_{e1}=n_{i1}$, if the electron and ion number densities are only perturbed.
However, in our general calculations in the Appendix, the values
$\mathbf{\nabla\cdot v}_{e1}$ and $\mathbf{\nabla\cdot v}_{i1}$ are considered
as different in the cases $n_{e0}\neq n_{i0}$ and $n_{e0}=n_{i0}$. Solving
equations of motion, we do not need to use the condition $n_{i1}=n_{e1}$.
These equations permit us to treat a general case $n_{i1}\neq n_{e1}$ that can
be appropriate for multicomponent plasmas. To derive the dispersion relation,
we find expressions for perturbed velocities and calculate the perturbed
current which then is used in Equation (8) in his given form. The perturbed
current does not contain density perturbations at the absent of background
flow velocities. Equations (24)-(29) are justified under conditions (18) and
(19) and can be applied for both $n_{i0}=n_{e0}$, $n_{i1}=n_{e1}$ and
$n_{i0}\neq n_{e0}$, $n_{i1}\neq n_{e1}$ (for multicomponent plasmas)\ cases.
Taking into account condition (18), we obtain from Equations (24), (25), (28),
and (29) that $\mathbf{\nabla\cdot v}_{i1}=\mathbf{\nabla\cdot v}%
_{e1}=-(c/B_{0})(\partial E_{1x}/\partial y)$ for perturbations (44).

From the results obtained above, we can estimate relative perturbations of
number density and pressure in the fast dynamical or long-wavelength
regime (18). Using, for example, Equations (2), (24) and (28) and keeping
main terms, we find equation for the ion density perturbation,%
\begin{equation}
\frac{\partial n_{i1}}{n_{i0}\partial t}=\frac{1}{\omega_{ci}}\frac{\partial
F_{i1x}}{\partial y}.
\end{equation}
From Equation (21), it follows that%

\begin{equation}
P_{i1}=\left(  -\lambda_{i}+\mu_{i}\right)  \frac{1}{\omega_{ci}}%
\frac{\partial^{2}F_{i1x}}{\partial y\partial t},
\end{equation}
where we have used the relation $m_{i}\mathbf{F}_{i1}=-m_{e}\mathbf{F}_{e1}$.
The value $P_{j1}$ is connected with the pressure perturbation $p_{j1}$ as
follows
\begin{equation}
P_{j1}=-\frac{1}{m_{j}n_{j0}}\frac{\partial p_{j1}}{\partial t}.
\end{equation}
From Equations (61)-(63), we obtain%
\begin{equation}
\frac{\partial n_{i1}}{n_{i0}\partial t}=\frac{T_{i0}}{\left(  \lambda_{i}%
-\mu_{i}\right)  m_{i}}\frac{p_{i1}}{p_{i0}}.
\end{equation}
Analogously, we have for electrons%
\begin{equation}
\frac{\partial n_{e1}}{n_{i0}\partial t}=\frac{T_{e0}}{\left(  \lambda_{e}%
-\mu_{e}\right)  m_{e}}\frac{p_{e1}}{p_{e0}},
\end{equation}
where $n_{e1}=n_{i1}$. Thus,%
\[
\frac{p_{i1}}{\left(  \lambda_{i}-\mu_{i}\right)  m_{i}}=\frac{p_{e1}}{\left(
\lambda_{e}-\mu_{e}\right)  m_{e}}.
\]
Taking into account notations (23), we see that $n_{i,e1}/n_{i0}\sim
p_{i,e1}/p_{e0}$ and $p_{i1}/p_{i0}\sim p_{e1}/p_{e0}$. Using the dispersion
relation (44) in the case of neglect the magnetic field, we obtain from
Equations (64) and (65) equation connecting the sum of pressures and density
perturbation,%
\[
\frac{\partial^{2}n_{i1}}{\partial t^{2}}=\frac{1}{m_{i}}\frac{\partial^{2}%
}{\partial y^{2}}\left(  p_{e1}+p_{i1}\right)  .
\]
It is easy to see from Equations (33), (38), and (44) that Equation (51) corresponds to isobaric regime where the sum of electron and ion pressure perturbations is smaller then electron or ion pressure perturbation (see also Nekrasov 2011). The perturbation of number density is due to electric drift (see Equation (61)).

Dispersion relation (50) is satisfied in the case $\omega^{2}\gg k_{y}%
^{2}c_{s}^{2}$. Taking into account condition (18), unstable perturbations
have $k_{y}^{2}\lesssim\left(  m_{i}/m_{e}\right)  k_{z}^{2}$. Thus, the
transverse wavelength $\lambda_{\perp}\gtrsim\left(  m_{e}/m_{i}\right)
^{1/2}\lambda_{z}$ and can be both less and larger than the longitudinal
wavelength $\lambda_{z}$. From other side, in the case $\omega^{2}\ll
k_{y}^{2}c_{s}^{2}$, dispersion relation (51) describes unstable perturbations
strongly elongated along the magnetic field, $k_{y}^{2}\gg\left(  m_{i}%
/m_{e}\right)  k_{z}^{2}$ or $\lambda_{z}\gg\left(  m_{i}/m_{e}\right)
^{1/2}\lambda_{\perp}$. In this case, very thin filaments are generated. Thus,
a wide spectrum of wavelengths of perturbations along and across the magnetic
field can be formed in the framework of conditions (18), (19), (34), and (41).

A general form of dispersion relation (46) including the thermal exchange
and different temperatures allows us to consider various cases, which can be
realized in real situations. We can investigate a weakly, strongly, and
intermediate thermal coupling (see Sections 5.5 and 5.6). In particular,
Equations (52) and (56) are available for a weak thermal coupling, while
Equations (53), (54), (57), and (58) are appropriate in the case of strong
coupling. The intermediate case is described by Equation (59), where
coefficients (60) contain both different temperatures and different cooling functions.

We have shown that unstable perturbations have an electromagnetic nature (see
Equation (44)). Thus, a consideration of only potential perturbations is in
general not adequate.

\bigskip

\section{\bigskip ASTROPHYSICAL\ IMPLICATIONS}

We shortly outline some important points of our investigation for possible
observations. We have found growth rates, which contain in the clear form
the separate contribution of cooling functions of electrons and ions. It
is obvious that both components (in multicomponent media also dust grains,
neutrals, and so on) can result in thermal instability in the same extent. This fact considerably extends possibilities for the medium to become unstable. In this
connection, it is important to know the functional dependence of cooling
functions on the temperature and density for each species. Unfortunately, at
present, there is not sufficient information on this subject in
astrophysical literature. The range of scale lengths of unstable
perturbations can enlarge due to contribution to instability of other species
except electrons. For example, short-wavelength perturbations, which must
be stable because of a large electron thermal conduction, can be unstable due
to contribution of ions to cooling of medium (see Section 9). In the
long-wavelength regime (18), scale sizes of unstable perturbations across
the magnetic field can have a wide spectrum and be, in particular, very
elongated along the magnetic field. Such filaments are observed in galaxy
clusters (e.g., Conselice et al. 2001; Salom\'{e} et al. 2006) and in the
solar corona (e.g., Tandberg-Hanssen 1974; Karpen et al. 1989). Different
temperatures of electrons and ions assumed in this paper can be observed in
galaxy clusters (Markevitch et al. 1996; Fox \& Loeb 1997; Ettori \&
Fabian 1998; Takizawa 1998). This fact proves that dynamical and statistical
processes could have timescales of the same order. It is clear that real
situations in astrophysical objects are much more complicated to be captured
by simplified theories. Knowledge of fundamental processes and more detailed
conditions from observations are very important for theoretical models and, in
particular, for further investigation of thermal instabilities.

\bigskip

\section{\bigskip CONCLUSION}

We have studied thermal instability in the electron-ion magnetized plasma
which is relevant to galaxy clusters, solar corona, and other two-component
astrophysical objects. The multicomponent plasma approach have been applied to
derive the dispersion relation for the condensation mode in the case in which
the dynamical time is smaller than a time the particles need to cover the
wavelength of perturbations along the magnetic field due to their thermal
velocity. Our dispersion relation takes into account the electron and ion
cooling-heating functions, collisions in momentum equations, energy
exchange in thermal equations, different background temperatures of electrons and ions, and perturbation of energy exchange collision frequency due to
density and temperature perturbations. Different limiting cases of
dispersion relation have been considered and simple expressions for growth
rates have been obtained. We have shown that perturbations have an
electromagnetic nature. We have found that at conditions under consideration
transverse scale sizes of unstable perturbations can have a wide spectrum
relatively to longitudinal scale sizes and, in particular, form very thin
filaments. General expressions for dynamical variables obtained in this
paper can be applied for astrophysical and laboratory plasmas also containing
the neutrals, dust grains, and other species. The results obtained can be useful for interpretation of observations of dense cold regions in astrophysical objects.

In this paper, we have investigated the linear stage of thermal instability in
the multicomponent medium. The instability development can result in plasma
turbulence when transport coefficients become dependent not on Coulomb
collisions but on the energy of turbulence. In this case, a nonlinear
consideration of a problem is necessary.

\bigskip

\section{ACKNOWLEDGMENTS}

I would like gratefully to acknowledge Mohsen Shadmehri for valuable
discussions and suggestions and anonymous referee for his/her constructive and detailed comments which have allowed to improve the manuscript.

\bigskip

\begin{appendix}
\section{APPENDIX}
\subsection{Perturbed velocities of species}
In the linear approximation, Equation (1) for the perturbed velocity
$\mathbf{v}_{j1}$ takes the form%
\begin{equation}
\frac{\partial\mathbf{v}_{j1}}{\partial t}=-\frac{\mathbf{\nabla}p_{j1}}%
{m_{j}n_{j0}}+\mathbf{F}_{j1}\mathbf{+}\frac{q_{j}}{m_{j}c}\mathbf{v}%
_{j1}\times\mathbf{B}_{0},
\end{equation}
where $p_{j1}=n_{j0}T_{j1}+n_{j1}T_{j0}$. From this equation, we can find
solutions for the components of $\mathbf{v}_{j1}$. For simplicity, we assume
that $\partial/\partial x=0$ because a system is symmetric in the transverse
direction relative to the $z$-axis. Then, the $x$-component of Equation (A1)
gives%
\begin{equation}
\frac{\partial v_{j1x}}{\partial t}=F_{j1x}\mathbf{+}\omega_{cj}v_{j1y},
\end{equation}
where $\omega_{cj}=q_{j}B_{0}/m_{j}c$ is the cyclotron frequency.
Differentiating Equation (A1) over $t$ and using Equation (2) in the linear
approximation and Equations (15), (16), and (A2), we obtain for the
$y$-component of Equation (A1)%
\begin{equation}
\left(  \frac{\partial^{2}}{\partial t^{2}}+\omega_{cj}^{2}\right)
v_{j1y}=\frac{\partial P_{j1}}{\partial y}+Q_{j1y},
\end{equation}
where%
\begin{align}
P_{e1}  & =-\frac{G_{2}}{Dm_{e}}\frac{\partial}{\partial t}\mathbf{\nabla
}\cdot\mathbf{v}_{i1}+\left(  \frac{T_{e0}}{m_{e}}-\frac{G_{1}}{Dm_{e}}%
\frac{\partial}{\partial t}\right)  \ \mathbf{\nabla}\cdot\mathbf{v}_{e1},\\
P_{i1}  & =-\frac{G_{3}}{Dm_{i}}\frac{\partial}{\partial t}\mathbf{\nabla
}\cdot\mathbf{v}_{e1}+\left(  \frac{T_{i0}}{m_{i}}-\frac{G_{4}}{Dm_{i}}%
\frac{\partial}{\partial t}\right)  \mathbf{\nabla}\cdot\mathbf{v}%
_{i1}.\nonumber
\end{align}
The value $P_{j1}$ is connected with the pressure perturbation (see Equation
(A1)). Using Equations (A2) and (A3), we find
\begin{equation}
\frac{\partial}{\omega_{cj}\partial t}\left[  \left(  \frac{\partial^{2}%
}{\partial t^{2}}+\omega_{cj}^{2}\right)  v_{j1x}-Q_{j1x}\right]
\mathbf{=}\frac{\partial P_{j1}}{\partial y}%
\end{equation}
In Equations (A3) and (A5), notations%
\begin{align}
Q_{j1y}  & =-\omega_{cj}F_{j1x}+\frac{\partial F_{j1y}}{\partial t},\\
Q_{j1x}  & =\omega_{cj}F_{j1y}\mathbf{+}\frac{\partial F_{j1x}}{\partial
t}\nonumber
\end{align}
are introduced. We see from these equations that the thermal pressure effect
on the velocity $v_{i1x}$ is much larger than that on $v_{i1y}$ when
$\partial/\partial t\ll\omega_{ci}$. The $z$-component of Equation (A1) can be
written in the form%
\begin{equation}
\frac{\partial^{2}v_{j1z}}{\partial t^{2}}=\frac{\partial P_{j1}}{\partial
z}+\frac{\partial F_{j1z}}{\partial t}.
\end{equation}
\subsection{Calculation of $\mathbf{\nabla\cdot v}_{j1}$ and $P_{j1} $}
We have%
\begin{equation}
\mathbf{\nabla}\cdot\mathbf{v}_{j1}=\frac{\partial v_{j1y}}{\partial y}%
+\frac{\partial v_{j1z}}{\partial z}.
\end{equation}
Using Equations (A3), (A4), (A7), and (A8), we obtain%
\begin{align}
L_{1e}\ \mathbf{\nabla}\cdot\mathbf{v}_{e1}+L_{2e}\mathbf{\nabla}%
\cdot\mathbf{v}_{i1}  & =H_{e1},\\
L_{1i}\mathbf{\nabla}\cdot\mathbf{v}_{i1}+L_{2i}\mathbf{\nabla}\cdot
\mathbf{v}_{e1}  & =H_{i1}.\nonumber
\end{align}
Here%
\begin{equation}
H_{j1}=\frac{\partial^{3}Q_{j1y}}{\partial y\partial t^{2}}+\left(
\frac{\partial^{2}}{\partial t^{2}}+\omega_{cj}^{2}\right)  \frac{\partial
^{2}F_{j1z}}{\partial z\partial t}%
\end{equation}
and operators $L_{1j}$ and $L_{2j}$ are the following:%
\begin{align}
L_{1e}  & =\left(  \frac{\partial^{2}}{\partial t^{2}}+\omega_{ce}^{2}\right)
\frac{\partial^{2}}{\partial t^{2}}-L_{3e}\left(  \frac{T_{e0}}{m_{e}}%
-\frac{G_{1}}{Dm_{e}}\frac{\partial}{\partial t}\right)  ,\\
L_{1i}  & =\left(  \frac{\partial^{2}}{\partial t^{2}}+\omega_{ci}^{2}\right)
\frac{\partial^{2}}{\partial t^{2}}-L_{3i}\left(  \frac{T_{i0}}{m_{i}}%
-\frac{G_{4}}{Dm_{i}}\frac{\partial}{\partial t}\right)  ,\nonumber\\
L_{2e}  & =L_{3e}\frac{G_{2}}{Dm_{e}}\frac{\partial}{\partial t},L_{2i}%
=L_{3i}\frac{G_{3}}{Dm_{i}}\frac{\partial}{\partial t},\nonumber\\
L_{3j}  & =\frac{\partial^{4}}{\partial y^{2}\partial t^{2}}\ +\left(
\frac{\partial^{2}}{\partial t^{2}}+\omega_{cj}^{2}\right)  \frac{\partial
^{2}}{\partial z^{2}}.\nonumber
\end{align}
From a system of equations (A9), we find
\begin{align}
L\mathbf{\nabla}\cdot\mathbf{v}_{e1}  & =-L_{2e}H_{i1}+L_{1i}H_{e1},\\
L\mathbf{\nabla}\cdot\mathbf{v}_{i1}  & =-L_{2i}H_{e1}+L_{1e}\ H_{i1}%
,\nonumber
\end{align}
where
\begin{equation}
L=L_{1e}L_{1i}\ -L_{2e}L_{2i}.
\end{equation}
The values $P_{e1}$ and $P_{i1}$ can be found, substituting solutions (A12)
into expressions (A4),%
\begin{align}
LP_{i1}  & =\left[  \frac{G_{3}}{Dm_{i}}\frac{\partial}{\partial t}%
L_{2e}+\left(  \frac{T_{i0}}{m_{i}}-\frac{G_{4}}{Dm_{i}}\frac{\partial
}{\partial t}\right)  L_{1e}\right]  \ H_{i1}\\
& -\left[  \frac{G_{3}}{Dm_{i}}\frac{\partial}{\partial t}L_{1i}+\left(
\frac{T_{i0}}{m_{i}}-\frac{G_{4}}{Dm_{i}}\frac{\partial}{\partial t}\right)
L_{2i}\right]  H_{e1}.\nonumber
\end{align}%
\begin{align}
LP_{e1}  & =\left[  \frac{G_{2}}{Dm_{e}}\frac{\partial}{\partial t}%
L_{2i}+\left(  \frac{T_{e0}}{m_{e}}-\frac{G_{1}}{Dm_{e}}\frac{\partial
}{\partial t}\right)  L_{1i}\right]  H_{e1}\\
& -\left[  \frac{G_{2}}{Dm_{e}}\frac{\partial}{\partial t}L_{1e}\ +\left(
\frac{T_{e0}}{m_{e}}-\frac{G_{1}}{Dm_{e}}\frac{\partial}{\partial t}\right)
L_{2e}\right]  H_{i1},\nonumber
\end{align}
\subsection{Expressions for $D$ and $G_{1,2,3,4}$}
We now give expressions for values defined by a system (17):
\begin{equation}
D=\left(  \frac{\partial}{\partial t}+\Omega_{\chi}+\Omega_{Te}\right)
\left(  \frac{\partial}{\partial t}+\Omega_{Ti}+\Omega_{ie}\right)
\frac{\partial^{2}}{\partial t^{2}}+\left(  \Omega_{ei}+\Omega_{Tei}\right)
\left(  \frac{\partial}{\partial t}+\Omega_{Ti}\right)  \frac{\partial^{2}%
}{\partial t^{2}},
\end{equation}%
\begin{equation}
G_{1}=T_{e0}\left[  \Omega_{ne}-\left(  \gamma-1\right)  \frac{\partial
}{\partial t}\right]  \left(  \frac{\partial}{\partial t}+\Omega_{Ti}%
+\Omega_{ie}\right)  \frac{\partial}{\partial t}+\Omega_{ei}\left(
T_{e0}-T_{i0}\right)  \left(  \frac{\partial}{\partial t}+\Omega_{Ti}\right)
\frac{\partial}{\partial t},
\end{equation}%
\begin{equation}
G_{2}=\Omega_{ei}T_{i0}\left[  \Omega_{ni}-\left(  \gamma-1\right)
\frac{\partial}{\partial t}\right]  \frac{\partial}{\partial t}+\Omega
_{ei}\left(  T_{e0}-T_{i0}\right)  \left(  \frac{\partial}{\partial t}%
+\Omega_{Ti}\right)  \frac{\partial}{\partial t},
\end{equation}%
\begin{equation}
G_{3}=\left(  \Omega_{Tie}+\Omega_{ie}\right)  T_{e0}\left[  \Omega
_{ne}-\left(  \gamma-1\right)  \frac{\partial}{\partial t}\right]
\frac{\partial}{\partial t}-\Omega_{ie}\left(  T_{e0}-T_{i0}\right)  \left(
\frac{\partial}{\partial t}+\Omega_{\chi}+\Omega_{Te}\right)  \frac{\partial
}{\partial t},
\end{equation}%
\begin{align}
G_{4}  & =T_{i0}\left(  \frac{\partial}{\partial t}+\Omega_{\chi}+\Omega
_{Te}+\Omega_{Tei}+\Omega_{ei}\right)  \left[  \Omega_{ni}-\left(
\gamma-1\right)  \frac{\partial}{\partial t}\right]  \frac{\partial}{\partial
t}\\
& -\Omega_{ie}\left(  T_{e0}-T_{i0}\right)  \left(  \frac{\partial}{\partial
t}+\Omega_{\chi}+\Omega_{Te}\right)  \frac{\partial}{\partial t}.\nonumber
\end{align}
\subsection{Simplification of Equations (A14) and (A15)}
We further calculate coefficients by $H_{j1}$ in Equations (A14) and
(A15). Using expressions (A11), we find
\begin{equation}
\frac{G_{3}}{Dm_{i}}\frac{\partial}{\partial t}L_{1i}+\left(  \frac{T_{i0}%
}{m_{i}}-\frac{G_{4}}{Dm_{i}}\frac{\partial}{\partial t}\right)  L_{2i}%
=\frac{G_{3}}{Dm_{i}}\left(  \frac{\partial^{2}}{\partial t^{2}}+\omega
_{ci}^{2}\right)  \frac{\partial^{3}}{\partial t^{3}}%
\end{equation}
and%
\begin{equation}
\frac{G_{3}}{Dm_{i}}\frac{\partial}{\partial t}L_{2e}+\left(  \frac{T_{i0}%
}{m_{i}}-\frac{G_{4}}{Dm_{i}}\frac{\partial}{\partial t}\right)  L_{1e}%
=\frac{1}{D}\left(  D\frac{T_{i0}}{m_{i}}-\frac{G_{4}}{m_{i}}\frac{\partial
}{\partial t}\right)  \left(  \frac{\partial^{2}}{\partial t^{2}}+\omega
_{ce}^{2}\right)  \frac{\partial^{2}}{\partial t^{2}}+\frac{1}{Dm_{e}m_{i}%
}L_{3e}K.
\end{equation}
In Equation (A22), we have introduced notation%
\begin{equation}
K=\frac{1}{D}\left(  G_{2}G_{3}-G_{1}G_{4}\right)  \frac{\partial^{2}%
}{\partial t^{2}}+\left(  T_{e0}G_{4}+T_{i0}G_{1}\right)  \frac{\partial
}{\partial t}-DT_{e0}T_{i0}.
\end{equation}
Calculations show that the value $\left(  G_{2}G_{3}-G_{1}G_{4}\right)  $ has
a simple form, i.e.,%
\begin{align}
\frac{1}{D}\left(  G_{2}G_{3}-G_{1}G_{4}\right)   & =\Omega_{ie}\left(
T_{e0}-T_{i0}\right)  T_{e0}\left[  \Omega_{ne}-\left(  \gamma-1\right)
\frac{\partial}{\partial t}\right]  +\Omega_{ei}\left(  T_{i0}-T_{e0}\right)
T_{i0}\left[  \Omega_{ni}-\left(  \gamma-1\right)  \frac{\partial}{\partial
t}\right] \\
& -T_{e0}T_{i0}\left[  \Omega_{ne}-\left(  \gamma-1\right)  \frac{\partial
}{\partial t}\right]  \left[  \Omega_{ni}-\left(  \gamma-1\right)
\frac{\partial}{\partial t}\right]  .\nonumber
\end{align}
Using expressions (A16), (A17), (A20), and (A24), we obtain for the operator
$K$ (A23) the simple form,
\begin{equation}
K=-\Omega_{ie}T_{e0}^{2}W_{e}\frac{\partial^{2}}{\partial t^{2}}-\left(
\Omega_{ei}T_{i0}+\Omega_{Tei}T_{e0}\right)  T_{i0}W_{i}\frac{\partial^{2}%
}{\partial t^{2}}-T_{e0}T_{i0}W_{e}W_{i}\frac{\partial^{2}}{\partial t^{2}},
\end{equation}
where notations
\begin{align}
W_{e}  & =\gamma\frac{\partial}{\partial t}+\Omega_{\chi}+\Omega_{Te}%
-\Omega_{ne},\\
W_{i}  & =\gamma\frac{\partial}{\partial t}+\Omega_{Ti}-\Omega_{ni}\nonumber
\end{align}
are introduced. Using Equations (A21) and (A22), Equation (A14) for $P_{i1}$
takes the form,
\begin{align}
DLP_{i1}  & =\left[  \left(  D\frac{T_{i0}}{m_{i}}-\frac{G_{4}}{m_{i}}%
\frac{\partial}{\partial t}\right)  \left(  \frac{\partial^{2}}{\partial
t^{2}}+\omega_{ce}^{2}\right)  \frac{\partial^{2}}{\partial t^{2}}+\frac
{1}{m_{e}m_{i}}L_{3e}K\right]  \ H_{i1}\\
& -\frac{G_{3}}{m_{i}}\left(  \frac{\partial^{2}}{\partial t^{2}}+\omega
_{ci}^{2}\right)  \frac{\partial^{3}}{\partial t^{3}}H_{e1}.\nonumber
\end{align}
Analogous consideration of Equation (A15) leads to the following equation for
$P_{e1}$:
\begin{align}
DLP_{e1}  & =\left[  \left(  D\frac{T_{e0}}{m_{e}}-\frac{G_{1}}{m_{e}}%
\frac{\partial}{\partial t}\right)  \left(  \frac{\partial^{2}}{\partial
t^{2}}+\omega_{ci}^{2}\right)  \frac{\partial^{2}}{\partial t^{2}}+\frac
{1}{m_{i}m_{e}}L_{3i}K\right]  H_{e1}\\
& -\frac{G_{2}}{m_{e}}\left(  \frac{\partial^{2}}{\partial t^{2}}+\omega
_{ce}^{2}\right)  \frac{\partial^{3}}{\partial t^{3}}H_{i1}.\nonumber
\end{align}
Operators
\begin{align*}
& D\frac{T_{e0}}{m_{e}}-\frac{G_{1}}{m_{e}}\frac{\partial}{\partial t},\\
& D\frac{T_{i0}}{m_{i}}-\frac{G_{4}}{m_{i}}\frac{\partial}{\partial t}%
\end{align*}
can be found by using Equations (A16), (A17), (A20), and (A26)%
\begin{align}
D\frac{T_{e0}}{m_{e}}-\frac{G_{1}}{m_{e}}\frac{\partial}{\partial t}  &
=\frac{T_{e0}}{m_{e}}W_{e}\left(  \frac{\partial}{\partial t}+\Omega
_{Ti}+\Omega_{ie}\right)  \frac{\partial^{2}}{\partial t^{2}}\\
& +\frac{1}{m_{e}}\left(  T_{e0}\Omega_{Tei}+\Omega_{ei}T_{i0}\right)  \left(
\frac{\partial}{\partial t}+\Omega_{Ti}\right)  \frac{\partial^{2}}{\partial
t^{2}},\nonumber\\
D\frac{T_{i0}}{m_{i}}-\frac{G_{4}}{m_{i}}\frac{\partial}{\partial t}  &
=\frac{T_{i0}}{m_{i}}W_{i}\left(  \frac{\partial}{\partial t}+\Omega_{\chi
}+\Omega_{Te}+\Omega_{ei}+\Omega_{Tei}\right)  \frac{\partial^{2}}{\partial
t^{2}}\nonumber\\
& +\frac{T_{e0}}{m_{i}}\Omega_{ie}\left(  \frac{\partial}{\partial t}%
+\Omega_{\chi}+\Omega_{Te}\right)  \frac{\partial^{2}}{\partial t^{2}%
}.\nonumber
\end{align}
\subsection{Operator $L$ in a general form}
Using expressions (A11), we find from Equation (A13)%
\begin{equation}
L=M-N-\frac{1}{m_{e}m_{i}D}L_{3e}L_{3i}K,
\end{equation}
where
\begin{align}
M  & =\left(  \frac{\partial^{2}}{\partial t^{2}}+\omega_{ce}^{2}\right)
\left(  \frac{\partial^{2}}{\partial t^{2}}+\omega_{ci}^{2}\right)
\frac{\partial^{4}}{\partial t^{4}},\\
N  & =\left(  \frac{\partial^{2}}{\partial t^{2}}+\omega_{ci}^{2}\right)
\frac{\partial^{2}}{\partial t^{2}}L_{3e}\left(  \frac{T_{e0}}{m_{e}}%
-\frac{G_{1}}{Dm_{e}}\frac{\partial}{\partial t}\right)  +\left(
\frac{\partial^{2}}{\partial t^{2}}+\omega_{ce}^{2}\right)  \frac{\partial
^{2}}{\partial t^{2}}L_{3i}\left(  \frac{T_{i0}}{m_{i}}-\frac{G_{4}}{Dm_{i}%
}\frac{\partial}{\partial t}\right)  ,\nonumber
\end{align}
and $K$ is defined by Equation (A25).
\end{appendix}

\bigskip

\section{\bigskip REFERENCES}

Audit, E., \& Hennebelle, P. 2005, A\&A, 433,1

Balbus, S. A. 1991, ApJ, 372, 25

Begelman, M. C., \& Zweibel, E. G. 1994, ApJ, 431, 689

Birk, G. T. 2000, Phys. Plasmas, 7, 3811

Birk, G. T., \& Wiechen, H. 2001, Phys. Plasmas, 8, 5057

Bogdanovi\'{c}, T., Reynolds, C. S., Balbus, S. A., \& Parrish, I. J. 2009,
ApJ, 704, 211

Braginskii, S. I. 1965, Rev. Plasma Phys., 1, 205

Burkert, A., \& Lin, D. N. C. 2000, ApJ, 537, 270

Conselice, C. J., Gallagher, J. S., III, \& Wyse, R. F. G. 2001, AJ, 122, 2281

Cox, D. P. 2005, ARA\&A, 43, 337

Elmegreen, B. G., \& Scalo, J. 2004, ARA\&A, 42, 211

Ettori, S. \& Fabian, A. C. 1998, MNRAS, 293, L33

Field, G.B. 1965, ApJ, 142, 531

Fox, D. C., \& Loeb, A. 1997, ApJ, 491, 459

Fukue, T., \& Kamaya, H. 2007, ApJ, 669, 363

Heiles, C., \& Crutcher, R. 2005, in Cosmic Magnetic Fields, ed. R.Wielebinski

\& R. Beck (Lecture Notes in Physics) (Berlin: Springer)

Hennebelle, P., \& P\'{e}rault, M. 2000, A\&A, 359, 1124

Heyvaerts, J. 1974, A\&A, 37, 65

Ib\'{a}\~{n}ez, M. H., \& Shchekinov, Yu. A. 2002, Phys. Plasmas, 9, 3259

Inoue, T., \& Inutsuka, S. 2008, ApJ, 687, 303

Karpen, J. T., Antiochos, S. K., Picone, J. M., \& Dahlburg, R. B. 1989, ApJ,
338, 493

Kopp, A., Schr\"{o}er, A., Birk, G. T., \& Shukla, P. K. 1997, Phys. Plasmas,
4, 4414

Kopp, A., \& Shchekinov, Yu. A. 2007, Phys. Plasmas, 14, 073701

Koyama, H., \& Inutsuka, S. 2002, ApJ, 564, L97

Kritsuk, A. G., \& Norman, M. L. 2002, ApJ, 569, L127

Loewenstein, M. 1990, ApJ, 349, 471

Markevitch M., Mushotzky R., Inoue H., Yamashita K., Furuzawa A.,

Tawara Y., 1996, ApJ, 456, 437

Mason, S. F., \& Bessey, R. J. 1983, Solar Phys., 83, 121

Meerson, B. 1996, Rev. Mod. Phys., 68, 215

Nakagawa, Y. 1970, Sol. Phys., 12, 419

Nekrasov, A. K. 2009a, ApJ, 695, 46

Nekrasov, A. K. 2009b, ApJ, 704, 80

Nekrasov, A. K. 2009c, MNRAS, 400, 1574

Nekrasov, A. K. 2011, ApJ (accepted)

Nekrasov, A. K., \& Shadmehri, M. 2010, ApJ, 724, 1165

Nekrasov, A. K., \& Shadmehri, M. 2011, Astrophys. Space Sci., 333, 477

Pandey, B. P., \& Krishan, V. 2001, IEEE Trans. Plasma Sci., 29, 307

Pandey, B. P., Vranje\v{s}, J., \& Parshi, S. 2003, Pramana, 60, 491

Parker, E. N. 1953, ApJ, 117, 431

Parrish, I. J., Quataert, E., \& Sharma, P. 2009, ApJ., 703, 96

Salom\'{e}, P., Combes, F., Edge, A. C., et al. 2006, A\&A, 454, 437

S\'{a}nchez-Salcedo, F. J., V\'{a}zquez-Semadeni, E., \& Gazol, A. 2002, ApJ,
577, 768

Shadmehri, M., Nejad-Asghar, M., \& Khesali, A. 2010, Ap\&SS, 326, 83

Sharma, P., Parrish, I. J., \& Quataert, E. 2010, ApJ, 720, 652

Shukla, P. K., \& Sandberg, I. 2003, Phys. Rev. E, 67, 036401

Stiele, H., Lesch, H., \& Heitsch, F. 2006, MNRAS, 372, 862

Takizawa, M. 1998, ApJ, 509, 579

Tandberg-Hanssen, E. 1974, Solar Prominences (Dordrecht, Holland: D. Reidel
Publ. Co.)

Tozzi, P., \& Norman, C. 2001, ApJ, 546, 63

Trevisan, M. C., \& Ib\'{a}\~{n}ez, M. H. 2000, Phys. Plasmas, 7, 897

V\'{a}zquez-Semadeni, E., Gazol, A., Passot, T., \& S\'{a}nchez-Salcedo, F. J. 2003, in Turbulence and Magnetic Fields in Astrophysics, ed. E. Falgarone \& T. Passot (Berlin: Springer), 213

V\'{a}zquez-Semadeni, E., Ryu, D., Passot, T., Gonz\'{a}lez, R. F., \& Gazol,
A. 2006, ApJ, 643, 245

Yatou, H., \& Toh, S. 2009, Phys. Rev. E, 79, 036314

\section{\bigskip}

\end{document}